\documentclass[twocolumn,preprintnumbers,amsmath,amssymb]{revtex4}
\usepackage{graphicx}% Include figure files
\usepackage{bm}% bold math

\begin{document}

\title{Quantum computation in semiconductor quantum dots of electron-spin asymmetric anisotropic exchange}

\author{Xiang Hao}

\author{Shiqun Zhu}
\altaffiliation{Corresponding author} \email{szhu@suda.edu.cn}

\affiliation{School of Physical Science and Technology, Suzhou
University, Suzhou, Jiangsu 215006, People's Republic of China}

\begin{abstract}

The universal quantum computation is obtained when there exists
asymmetric anisotropic exchange between electron spins in coupled
semiconductor quantum dots. The asymmetric Heisenberg model can be
transformed into the isotropic model through the control of two
local unitary rotations for the realization of essential quantum
gates. The rotations on each qubit are symmetrical and depend on the
strength and orientation of asymmetric exchange. The implementation
of the axially symmetric local magnetic fields can assist the
construction of quantum logic gates in anisotropic coupled quantum
dots. This proposal can efficiently use each physical electron spin
as a logical qubit in the universal quantum computation.

\end{abstract}

\maketitle

\section{Introduction}

The electron or nuclear spins in quantum dots are usually regarded
as important candidates for qubits in quantum information because of
their long decoherence times \cite{Divin98, Vrijen00, Kane98}. It is
shown that any universal quantum computation can be implemented by a
series of arbitrary single-qubit rotations and controlled-NOT gate
\cite{Nielsen00}. The recent experiment demonstrates the feasibility
of coherent control of single-qubit gates by the short bursts of
oscillating magnetic fields \cite{Koppens06}. In many spin-based
quantum computation schemes, isotropic Heisenberg interactions are
dominant since the symmetric Hamiltonian conserves the total spin.
However, the general spin-orbit couplings actually exist in quantum
dots. The anisotropic interaction between electrons in conduction
band is very typical in solids
\cite{Kavokin01,Gorkov03,Kavokin04,Badescu05}. This coupling can
show the Dresselhaus form for the bulk inversion asymmetry
\cite{Dresselhaus55}. The heterostructure asymmetry can also induce
the Rashba coupling \cite{Levitov01}. These two couplings can bring
out the asymmetric anisotropic Dzyaloshinskii-Moriya (DM)
interaction \cite{Dzyaloshinskii58,Moriya60}. Unlike the isotropic
exchange, the asymmetric spin-orbit coupling can greatly reduce the
gate fidelity which cannot be neglected in quantum information
processing
\cite{Bonesteel01,Burkard02,Lida02,Stepanenko03,Stepanenko04}. To
eliminate the impacts of these asymmetric anisotropic couplings,
some useful methods of universal quantum computation were used by
means of encoding a logical qubit into more physical spins
\cite{DiVin00,Benjamin01,Lidar022,Levy02}. These encoding schemes
will unavoidably waste many resources of physical quantum spins.

In this paper, a method based on the control of symmetric
single-qubit rotations in semiconductor quantum dots is proposed.
The Hamiltonian of coupled electron spins with asymmetric
anisotropic exchange and the transformation method are presented in
Sec. II. In Sec. III, the perfect two-qubit gates are realized by
the local unitary operations in the new Hilbert spin space. A
discussion concludes the paper.

\section{Asymmetric Anisotropic Spin Coupling Model}

The recent work presented the microscopic description of the
interaction between electron spins in two coupled semiconductor
quantum dots \cite{Badescu05}. Different from the exchange of
electrons in conduction band, the asymmetric coupling arises from
the Coulomb interaction and from the conduction-valance band mixing.
In the twisted spin representation \cite{Levitov01}, the Hamiltonian
of the two asymmetric coupled spins $\vec{S}_{1}$ and $\vec{S}_{2}$
can be expressed by \cite{Badescu05}
\begin{equation}
\label{eq1}
H=H_{S}+H_{A}+H_{DM}
\end{equation}
where $H_{S}=J\cos \omega \vec{S}_1 \cdot \vec{S}_2$ represents the
symmetric isotropic Heisenberg interaction. The term of $H_A=2J(\sin
\frac {\omega}{2})^2(\vec{n} \cdot \vec{S}_1)(\vec{n} \cdot
\vec{S}_2)$ denotes the symmetric anisotropic interaction. The last
term $H_{DM}$ is the main part of asymmetric anisotropic exchange in
the Dzyaloshinskii-Moriya form of $H_{DM}=J\sin \omega [\vec{n}
\cdot (\vec{S}_1 \times \vec{S}_2)]$. The parameter $J$ is the
isotropic Heisenberg exchange constant, $\vec{n}=\vec{b}/|\vec{b}|$
is the orientation of the asymmetric anisotropic exchange $\vec{b}$.
The inherent parameter $\omega=\arctan (|\vec{b}|/J)$. The
asymmetric anisotropic exchange $\vec{b}$ depends on the separation
distance and the orientation between two coupled quantum dots
\cite{Badescu05}. It is found that the error of the gate from the
asymmetric anisotropic exchange $\vec{b}$ is far beyond the limit of
fault tolerant quantum computation $10^{-6}$
\cite{Shor95,Gottesman98,Kempe01}. Therefore, it is necessary to
investigate the method to eliminate the effects of the asymmetric
exchange.

By diagonalization of the Hamiltonian $H$, the asymmetric form can
be rotated into the isotropic one $H_{0}$ in the new spin Hilbert
space with
\begin{equation}
\label{eq2}
H_{0}=THT^{\dag}
\end{equation}
where $T$ is the unitary rotation. The new spin space can be
expanded by $\{T|00\rangle, T|01\rangle, T|10\rangle, T|11\rangle
\}$ where $|i\rangle,(i=0,1)$ is the single-qubit basis of spin
operator $S^z$ with the corresponding eigenvalues $\pm \hbar/2$. The
definite form of unitary rotation $T$ takes a key part in the
following scheme of quantum computation. Without losing the
generality, two typical kinds of asymmetric exchange orientations of
$\vec{n}_{xy}=\cos \theta \vec{e}_x+\sin \theta \vec{e}_y$ and
$\vec{n}_z=\vec{e}_z$ are considered.

In the case of $\vec{n}_{xy}$, the four eigenstates of the
Hamiltonian $H$ can be written as
\begin{widetext}
\begin{eqnarray}\label{eq3}
|\varphi_1\rangle &=& |\psi^{+}\rangle \\ \nonumber
|\varphi_2\rangle &=& \frac 1{\sqrt 2}[(\sin \theta +\cos \theta
\cos \frac {\omega}{2})|\phi^{-}\rangle+i(\cos \theta -\sin \theta
\cos \frac {\omega}{2})|\phi^{+}\rangle-i\sin \frac
{\omega}{2}|\psi^{-}\rangle] \\ \nonumber |\varphi_3\rangle &=&
\frac 1{\sqrt 2}[(\cos \theta +\sin \theta \cos \frac
{\omega}{2})|\phi^{+}\rangle-i(\sin\theta -\cos \theta \cos \frac
{\omega}{2})|\phi^{-}\rangle+\sin \frac {\omega}{2}|\psi^{-}\rangle] \\
\nonumber |\varphi_4\rangle &=& -i\cos \theta \sin \frac
{\omega}{2}|\phi^{-}\rangle-\sin \theta \sin \frac {\omega}{2}
|\phi^{+}\rangle+\cos \frac{\omega}{2}|\psi^{-}\rangle
\end{eqnarray}
\end{widetext}
Here the states $|\psi^{\pm}\rangle=\frac 1{\sqrt2}(|01\rangle \pm
|10\rangle)$ and $|\phi^{\pm}\rangle=\frac 1{\sqrt2}(|00\rangle \pm
|11\rangle)$ are the four Bell states in the space of $\{|00\rangle,
|01\rangle, |10\rangle, |11\rangle\}$. Meanwhile, the four Bell
states are the eigenstates of the isotropic Heisenberg exchange
Hamiltonian $H_0$. By means of the representation transformation,
the unitary rotation $T_{xy}$ can be given in the space of
$\{|00\rangle, |01\rangle, |10\rangle, |11\rangle \}$ with
\begin{widetext}
\begin{equation}\label{eq4}
T_{xy}(\omega,\theta)= \left(\begin{array}{cccc}
(\cos \frac {\omega}{4})^2e^{i(\theta-\frac {\pi}{4})}&\cos \frac {\omega}{4}\sin \frac {\omega}{4}e^{i\frac {\pi}{4}}&-\cos \frac {\omega}{4}\sin \frac {\omega}{4}e^{i\frac {\pi}{4}}&(\sin \frac {\omega}{4})^2e^{-i(\theta+\frac {\pi}{4})}\\
\cos \frac {\omega}{4}\sin \frac {\omega}{4}e^{i(\theta+\frac {\pi}{2})}&(\cos \frac {\omega}{4})^2&(\sin \frac {\omega}{4})^2&\cos \frac {\omega}{4}\sin \frac {\omega}{4}e^{-i(\theta+\frac {\pi}{2})}\\
\cos \frac {\omega}{4}\sin \frac {\omega}{4}e^{i(\theta-\frac {\pi}{2})}&(\sin \frac {\omega}{4})^2&(\cos \frac {\omega}{4})^2&\cos \frac {\omega}{4}\sin \frac {\omega}{4}e^{-i(\theta-\frac {\pi}{2})}\\
(\sin \frac {\omega}{4})^2e^{i(\theta+\frac {\pi}{4})}&\cos \frac
{\omega}{4}\sin \frac {\omega}{4}e^{-i\frac {\pi}{4}}&-\cos \frac
{\omega}{4}\sin \frac {\omega}{4}e^{-i\frac {\pi}{4}}&(\cos \frac
{\omega}{4})^2e^{-i(\theta-\frac {\pi}{4})}
\end{array}\right)
\end{equation}
\end{widetext}

Similarly, the four eigenstates of the Hamiltonian with the
orientation $\vec{n}_z$ can be expressed as
\begin{eqnarray}
\label{eq5}
|\varphi_1\rangle &=& |\phi^{-}\rangle \\ \nonumber
|\varphi_2\rangle &=& |\phi^{+}\rangle \\
\nonumber |\varphi_3\rangle &=& \frac 1{\sqrt{1+(\tan \frac
{\omega}{2})^2}}(|\psi^{+}\rangle+i\tan \frac {\omega}{2}|\psi^{-}\rangle) \\
\nonumber |\varphi_4\rangle &=& \frac 1{\sqrt{1+(\tan \frac
{\omega}{2})^2}}(|\psi^{-}\rangle+i\tan \frac
{\omega}{2}|\psi^{+}\rangle)
\end{eqnarray}
The definite form of the unitary operation $T_z$ in the space of
$\{|00\rangle, |01\rangle, |10\rangle, |11\rangle \}$ can also be
obtained by the representation transformation
\begin{equation}
\label{eq6}
T_{z}(\omega,\theta)= \left(\begin{array}{cccc} 1 & 0 & 0 & 0\\
0& e^{-i\frac {\omega}{2}} & 0 & 0 \\
0 & 0 & e^{i\frac {\omega}{2}} & 0 \\
0 & 0 & 0 & 1
\end{array}\right)
\end{equation}

It is clear that the unitary operations $T$ depend only on the
orientation of asymmetric anisotropic exchange $\vec{n}$ and the
strength ratio of the asymmetric exchange to the isotropic constant
$|\vec{b}|/J$. This means that one kind of inherent quantum-dots
structures can determine the corresponding kind of unitary operation
form $T$. After the operation of $T$, the asymmetric anisotropic
term will be eliminated in the Hamiltonian. In the new spin space,
the new form of the Hamiltonian takes on the isotropic Heisenberg
interaction of $H_{0}=J\vec{S'}_1 \cdot \vec{S'}_2$ where the new
spin operator can be expressed by $\vec{S'}=T\vec{S}T^{\dag}$. The
method to apply the unitary operation is one essential aspect in the
quantum-dots quantum computation.

\section{Realization of Universal Quantum Logical Gates}

Before further discussions, the major property of the unitary
operation $T$ needs to be investigated. It is found that the
entanglement of the thermal state $\rho(H)$ keeps the same as that
of $\rho(H_{0})$. Since the entanglement of the states cannot be
varied when the states are transformed by unitary local rotations,
the two-qubit operation $T$ can be expressed by the direct product
of two unitary local rotations $T=U_{1} \otimes U_{2}$. Recent
experiment implies that arbitrary single-qubit rotations can be
implemented in quantum dots \cite{Koppens06}. After the operation of
$T^{\dag}$, the two-qubit evolution operator can be expressed by
\begin{equation}
\label{eq7} \mathcal{U}(t)=\exp\{ -i\int_0^t H(t')dt' \}T^{\dag}
\end{equation}
When the evolution operator $\mathcal{U}(t)$ is applied to the
initial quantum state $|\Psi(0)\rangle$ for a period of time $\tau$,
the intermediate state can be written by
$\mathcal{U}(\tau)|\Psi(0)\rangle$. If the unitary rotation $T$ is
then employed, the whole quantum state can be obtained by
\begin{equation}
\label{eq8}
|\Psi(t)\rangle=T\exp\{ -i\int_0^t H(t')dt'
\}T^{\dag}|\Psi(0)\rangle
\end{equation}
Meanwhile, Eq. (8) can also be expressed as $|\Psi(t)\rangle=\exp\{
-i\int_0^tT H(t')T^{\dag}dt'\}|\Psi(0)\rangle=\exp \{-i
\int_0^tH_0(t')dt'\}$. Thus, when the evolution time satisfies
$\int_0^{\tau_s}Jdt'=J\tau_s=\pi(mod$ $2\pi)$, one can obtain the
swap quantum gate $\mathcal{U}_{sw}$ which just exchanges the
quantum states between two electron spins \cite{Divin98}. Moreover,
with the help of single-qubit gate sequence, the controlled-NOT gate
can be constructed by $e^{i\frac {\pi}{2}S_1^z}e^{-i\frac
{\pi}{2}S_2^z}\mathcal{U}^{1/2}_{sw}e^{i \pi
S_1^z}\mathcal{U}^{1/2}_{sw}$ where the square-root of the swap gate
$\mathcal{U}^{1/2}_{sw}$ corresponds to the half evolution time
$\frac {\tau_s}{2}$ of the operator $\mathcal{U}$.

It is important how to apply this unitary transformation $T$. It is
known that the operation $T$ can be decomposed by two local unitary
rotations $U_1$ and $U_2$. Arbitrary unitary operation at qubit $i$
can be written as \cite{Barenco95}
\begin{equation}
\label{eq9}
U_i=\Phi_i(\delta)R_i^z(\alpha)R_i^y(\gamma)R_i^z(\beta)
\end{equation}
where $\Phi_i(\delta)=e^{i\delta I}$ is the phase shift with respect
to the angle $\delta$, $R_i^z(\alpha)=e^{i\alpha S_i^z}$ represents
the rotation by the angle $\alpha$ about the $z$-axis and
$R_i^y(\beta)=e^{i\beta S_i^y}$ is that by the angle $\beta$ about
the $y$-axis. It is found that the unitary rotation $T$ can be given
by single-qubit rotations as a function of the orientation and the
relative strength of the asymmetric exchange. For the two cases of
$\vec{n}_{xy}$ and $\vec{n}_{z}$, one has
\begin{eqnarray}\label{eq10}
T_{xy} &=& R_1^z(-\frac {3\pi}{4})R_1^y(\frac
{\omega}{2})R_1^z(\theta+\frac {\pi}{2}) \\ \nonumber & \otimes&
R_1^z(\frac
{\pi}{4})R_1^y(\frac {\omega}{2})R_1^z(\theta-\frac {\pi}{2}) \\
\nonumber T_{z} &=& R_1^z(-\frac {\omega}{2} ) \otimes R_2^z(\frac
{\omega}{2})
\end{eqnarray}
This equation demonstrates that the special unitary operation $T$ is
accomplished by two local axis-symmetrical rotations. Therefore, if
the unitary rotation $T$ is operated in the sequence of
$THT^{\dag}$, the asymmetric anisotropic exchange can be rotated
into the isotropic Heisenberg model with the same coupling $J$.

From Eq. (10), It is clear that these local rotations depend
critically on the values of the parameters $\omega$ and $\theta$.
The fidelity of one gate can be expressed by
\begin{equation}
\label{eq11} F=Tr[U^{\dag}U_{0}]/Tr[U_{0}^{\dag}U_{0}]
\end{equation}
where $U_0(\omega_0,\theta_0)$ is the perfect operation without
variations and $U(\omega,\theta)$ is the one with small variations
of $\Delta\omega=\omega-\omega_0$ and
$\Delta\theta=\theta-\theta_0$. The fidelity $F$ of swap gate
$\mathcal{U}_{sw}$ and the gate errors $\varepsilon$ are calculated
and plotted in Fig. 1 in dimensionless unit. The fidelity $F$ is
plotted in Fig. 1(a). It is seen that the gate fidelity decreases
when the variations $|\Delta\omega|/\omega_0$ and
$|\Delta\theta|/\theta_0$ increase. The gate error $\varepsilon=1-F$
is shown in Fig. 1(b). The order of the gate errors induced by the
asymmetric anisotropic exchange is about $5\times 10^{-6}\rightarrow
10^{-5}$ which is much higher than the limit of fault tolerant
quantum computation $10^{-6}$. Even when $|\Delta\theta|/\theta_0$
is varied from $0.1$ to $0.01$, the errors cannot be reduced and the
curves cannot be distinguished. By means of local single-qubit
operations $T_{xy}(\omega_{0}, \theta_{0})$ given by Eq. (10), the
order of gate errors can be reduced to about $10^{-7}$ which is much
smaller than $10^{-6}$. The error is reduced from $5\times
10^{-7}\rightarrow 10^{-8}$ when $|\Delta\theta|/\theta_0$ is varied
from $0.1$ to $0.01$. This means that the method in this paper is
feasible with the small variations of the asymmetric anisotropic
exchanges.

If there are external magnetic fields, the general Hamiltonian in
the inhomogeneous fields can be expressed as
\begin{equation}
\label{eq12}
H_{B}=H+(\vec{B}_1 \cdot \vec{S}_1+\vec{B}_2 \cdot \vec{S}_2)
\end{equation}
It is known that the isotropic Heisenberg interaction in the uniform
magnetic field can be regarded as one useful model for the universal
quantum computation \cite{DasSarma01}. The Hamiltonian $H$ in Eq.
(12) can be rotated into the isotropic one by the operation of
$THT^{\dag}$. The condition of what kind of the inhomogeneous fields
can be transformed in the same sequence of $T(\vec{B}_1 \cdot
\vec{S}_1+\vec{B}_2 \cdot \vec{S}_2)T^{\dag}=B(S_1^z+S_2^z)$ needs
to be investigated. It is demonstrated the axially symmetric
magnetic fields with the same strength satisfy this condition. For
the case of $\vec{n}_{xy}$, the previous inhomogeneous fields on two
qubits need to be applied in the form of
\begin{eqnarray}\label{eq13}
&\vec{B}_1&=-B\sin \frac {\omega}{2}(\sin \theta \vec{e}_x-\cos \theta \vec{e}_y-\cot \frac {\omega}{2}\vec{e}_z) \\
\nonumber &\vec{B}_2&=B\sin \frac {\omega}{2}(\sin \theta
\vec{e}_x-\cos \theta \vec{e}_y+\cot \frac {\omega}{2}\vec{e}_z)
\end{eqnarray}
Similarly, when the orientation of the asymmetric exchange
$\vec{n}_z=\vec{e}_z$, the magnetic fields on two qubits satisfy
$\vec{B}_1=\vec{B}_2=B\vec{e}_z$. Therefore, after the certain
period of time $\frac {\tau_s}{2}$, the phase-shifted swap action
can be constructed
\begin{eqnarray}\label{eq14}
&\mathcal{U}_{psw}&(a_1|0\rangle+b_1|1\rangle)_i \otimes
(a_2|0\rangle+b_2|1\rangle)_{i+1} \\ \nonumber &=&(a_2e^{\frac 12
g\mu_BB\tau_s}|0\rangle+b_2|1\rangle)_i\otimes
(a_1|0\rangle+b_1|1\rangle)_{i+1}
\end{eqnarray}
The controlled-NOT gate can be realized by the phase-shifted swap
gate and single-qubit rotations \cite{DasSarma01}.

\section{Discussion}

It is found that the asymmetric interactions can be transformed into
the isotropic Heisenberg model after the operation of local
symmetric rotations $U_1 \otimes U_2$. The local operations on each
qubit are determined by the orientation and the relative strength of
asymmetric exchange to the isotropic constant. The gate fidelity is
decreased when the small variations of the asymmetric anisotropy
increase. When the asymmetric anisotropy is eliminated by the
transformation, the order of gate errors is reduced to about
$10^{-8}$ which is much smaller than the limit of fault tolerant
quantum computation. Some special inhomogeneous magnetic fields can
also contribute to the realization of quantum logic gates. The
proposal can utilize each physical electron spin as a logical qubit
without encoding many spins.

{\bf Acknowledgement}

It is a pleasure to thank Yinsheng Ling, Jianxing Fang and Qing
Jiang for their many fruitful discussions about the topic. The
financial supports from the Specialized Research Fund for the
Doctoral Program of Higher Education of China (Grant No.
20050285002) and the National Natural Science Foundation of China
(Grant No. 10774108) are gratefully acknowledged.

\newpage

{\Large Fig. 1} (a). The fidelity of swap gate is plotted with small
variations of the asymmetric anisotropy when
$\tan\omega_0=5\times10^{-3}$ \cite{Badescu05} and $\theta_0=5\pi/6$
radius; (b). The gate error $\log_{10}\varepsilon$ is plotted when
$\Delta \theta/\theta_0=0.1, 0.01$. The dotted line denotes the
limit of fault tolerate quantum computation. The two solid lines
above the dotted line are the gate errors induced by $\vec{b}$.
However, they are not distinguishable. The two lines below the
dotted line are the errors induced by performing unitary local
rotations $T_{xy}(\omega_{0}, \theta_{0})$ with $\Delta
\theta/\theta_0=0.1, 0.01$ (from upper to lower).


\begin{references}

\bibitem{Divin98} D. Loss and D. P. DiVincenzo, Phys. Rev. A\textbf{57}, 120(1998).
\bibitem{Vrijen00} R. Vrijen, E. Yablonovitch, K. Wang, H. W. Jiang, A. Balandin,
V. Roychowdhury, T. Mor, and D. P. DiVincenzo, Phys. Rev. A\textbf{62}, 012306(2000).
\bibitem{Kane98} B. E. Kane, Nature \textbf{393}, 133(1998).
\bibitem{Nielsen00} M. A. Nielsen and I. L. Chuang, \textit{Quantum Computation and Quantum
Information}(Cambridge University Press, Cambridge, UK, 2000).
\bibitem{Koppens06} F. H. L. Koppens, C. Buizert, K. J. Tielrooij,
I. T. Vink, K. C. Nowack, T. Meunier, L. P. Kouwenhoven, and L. M.
K. Vandersypen, Nature \textbf{442}, 766(2006).
\bibitem{Kavokin01} K. V. Kavokin, Phys. Rev. B\textbf{64},
075305(2001).
\bibitem{Gorkov03} L. P. Gor'kov and P. L. Krotkov, Phys. Rev.
B\textbf{68}, 155206(2003).
\bibitem{Kavokin04} K. V. Kavokin, Phys. Rev. B\textbf{69},
075302(2004).
\bibitem{Badescu05} S. C. B\v{a}descu, Y. B. Lyanda-Geller, and T. L.
Reinecke, Phys. Rev. B\textbf{72}, 161304(R) (2005).
\bibitem{Dresselhaus55} G. Dresselhaus, Phys. Rev. \textbf{100},
580(1955).
\bibitem{Levitov01} L. S. Levitov and E. I. Rashba, Phys. Rev.
B\textbf{67}, 115324(2003).
\bibitem{Dzyaloshinskii58} I. Dzyaloshinskii, Phys. Chem. Solids
\textbf{4}, 241(1958).
\bibitem{Moriya60} T. Moriya, Phys. Rev. \textbf{120}, 91(1960).
\bibitem{Bonesteel01} N. E. Bonesteel, D. Stepanenko, and D. P.
DiVincenzo, Phys. Rev. Lett. \textbf{87}, 207901(2001).
\bibitem{Burkard02} G. Burkard and D. Loss, Phys. Rev. Lett.
\textbf{88}, 047903(2002).
\bibitem{Lida02} L. A. Wu and D. A. Lidar, Phys. Rev. A\textbf{66},
062314(2002).
\bibitem{Stepanenko03} D. Stepanenko, N. E. Bonesteel, D. P.
DiVincenzo, G. Burkard, and D. Loss, Phys. Rev. B\textbf{68},
115306(2003).
\bibitem{Stepanenko04} D. Stepanenko and N. E. Bonesteel, Phys. Rev.
Lett. \textbf{93}, 140501(2004).
\bibitem{DiVin00} D. P. DiVinzenco et. al. , Nature \textbf{408},
339(2000).
\bibitem{Benjamin01} S. C. Benjamin, Phys. Rev. A\textbf{64},
054303(2001).
\bibitem{Lidar022} D. A. Lidar and L. A. Wu, Phys. Rev.
Lett. \textbf{88}, 017905(2001)
\bibitem{Levy02} J. Levy, Phys. Rev. Lett. \textbf{89},
147902(2002).
\bibitem{Shor95} P. W. Shor, Phys. Rev. A\textbf{52},
R2493(1995).
\bibitem{Gottesman98} D. Gottesman, Phys. Rev. A\textbf{57},
127(1998).
\bibitem{Kempe01} J. Kempe, D. Bacon, D. A. Lidar, and K. B. Whaley,
Phys. Rev. A\textbf{63}, 042307(2001).
\bibitem{Barenco95} A. Barenco, C. H. Bennett, R. Cleve, D. P.
DiVincenzo, N. Margolus, P. Shor, T. Sleator, J. A. Smolin, and H.
Weinfurter Phys. Rev. A\textbf{52}, 3457(1995).
\bibitem{DasSarma01}X. Hu, R. de Sousa, and S. Das Sarma, Phys. Rev. Lett. \textbf{86}, 918(2001).

\end{references}
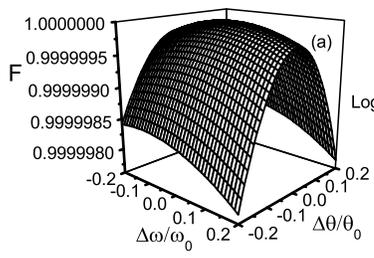
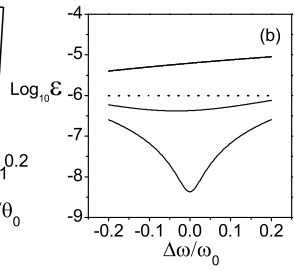
\end{document}